\documentclass[prd,onecolumn,nobibnotes,nofootinbib,showpacs]{revtex4}

\usepackage{amsmath,amssymb}
\usepackage{epsfig}
\usepackage{graphicx}

\newcommand{\med}[1]{\langle #1 \rangle}
\newcommand{\pslash}{\not{\hbox{\kern-2.pt $p$}}}
\newcommand{\kslash}{\not{\hbox{\kern-1.5pt $k$}}}
\newcommand{\qslash}{\not{\hbox{\kern-1.5pt $q$}}}
\newcommand{\lslash}{\not{\hbox{\kern-.3pt $l$}}}
\newcommand{\llslash}{\not{\hbox{\kern-.3pt $l_{1}$}}}
\newcommand{\lllslash}{\not{\hbox{\kern-.3pt $l_{2}$}}}

\begin{document}
\title{\bf Diffractive Higgs boson photoproduction in $\gamma p$ process}

\author{M. B. Gay Ducati$^{a,b}$}
\author{G. G. Silveira$^a$}

\affiliation{$^a$High Energy Physics Phenomenology Group, UFRGS, \\ Caixa Postal 15051, CEP 91501-970 - Porto Alegre, RS, Brazil.}

\affiliation{$^b$Centro Brasileiro de Pesquisas F\'{\i}sicas, Rua Dr.~Xavier Sigaud,~150, \\ CEP 22290-180 - Rio de Janeiro, RJ, Brazil.}

\begin{abstract}
We explore an alternative process for the diffractive Higgs boson production in peripheral pp collisions arising from Double Pomeron Exchange in photon-proton interaction. We introduce the impact factor formalism in order to enable the gluon ladder exchange in the photon-proton subprocess, and to permit the central Higgs production. The event rate for the diffractive Higgs production in central rapidity is estimated to be about 0.6 pb at Tevatron and LHC energies. This result is higher than predictions from other approaches for diffractive Higgs production, showing that the alternative production process leads to an enhanced signal for the detection of the Higgs boson at hadron colliders. Our results are compared to those obtained from a similar approach proposed by the Durham group. In this way we may examine the future developments in its application to pp and AA collisions.

\end{abstract}


\maketitle

\section{INTRODUCTION}

A new way to produce the Higgs boson in Peripheral Collisions at Hadron Colliders is calculated assuming an interaction through Double Pomeron Exchange (DPE) \cite{KMR-1997}. In $pp$ collisions, the interaction will occur between the colliding proton and the photon emitted from the electromagnetic field around the second proton \cite{bauretal,bertu,hencken}. Thus, the only way for an interaction to occur by DPE in a photon-proton process is to consider the photon splitting into a quark-antiquark pair, which enables one to use the impact factor formalism \cite{forshaw-book}. Adopting this mechanism for an elastic process, the final state of the exclusive event will be characterized by the presence of rapidity gaps between the proton and the Higgs, and between the photon and the Higgs.

Considering Peripheral Collisions, the gluons in the DPE will be exchanged in the $t$-channel of the photon-proton subprocess instead of the proton-proton system, allowing the impact factor formalism to be used to describe the splitting of the photon into a color dipole. In this model is convenient to consider null momentum transfer ($t = 0$) for the photon impact factor during the collision. Taking the Higgs mass as a hard scale, it is possible to safely compute the event rate in a perturbative way based on the Vector meson Dominance Model (VDM) \cite{barone}.

The plan of this paper is as follows. In Section \ref{sec:amp}, the scattering amplitude is calculated for the partonic $\gamma^{*}q$ subprocess with the photon virtuality applied for a quasi-real photon (Q$^{2} \simeq$ 0). The Section \ref{sec:proton} will be dedicated to analyse the process in a realistic way, where the quark contribution in the scattering amplitude will be replaced by a non-diagonal and non-integrated gluon distribution function in the proton of the exclusive $\gamma p$ process. The Section \ref{sec:results} provides the numerical results for the diffractive Higgs boson production, and we analyse the robustness of this approach. Finally, the Section \ref{sec:disc} discusses the important features of this approach and the Section \ref{sec:ccl} summarizes the ideas concerning the study of this physical process and the conclusions of this work.

\section{PARTONIC PROCESS}\label{sec:amp}

The study of diffractive Higgs boson production in $\gamma^{*}q$ processes is based on the kinematical variables used in the description of the Deeply Virtual Compton Scattering (DVCS), where the splitted photon interacts with the proton by a gluon ladder exchange \cite{frankfurt,kuchina}. The interaction between the colliding particles through DPE is the main feature of this proposal, which provides the leading process for the Higgs production in the range $M_{H} < $ 200 GeV. This kind of process is commonly studied in Peripheral Collisions, where the impact parameter between the colliding particles (protons or nuclei) is larger than their diameter. The protons only interact through the electromagnetic force \cite{bauretal}, enabling the $\gamma p$ or $\gamma \gamma$ subprocesses, where the photons are described by a Weizs\"acker-Williams distribution for each proton. Thus, in Peripheral Collisions the photons are treated as quasi-real particles due to the softness of its momentum.

\begin{figure}[t!]
\includegraphics[bb = 132 497 458 739 , scale=00.65]{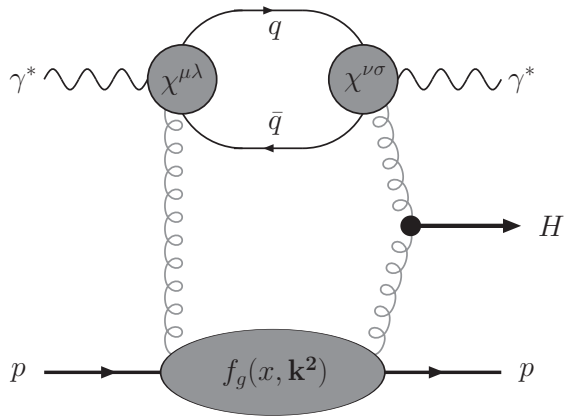}
\caption{\label{fig:foto-part} Feynman diagram representing the $\gamma^{*}p$ subprocess. The upper blobs represent the photon-gluon coupling which generate the color dipole. The momentum flux in the diagram is built in a way that the photon has no momentum transfer during its interaction with the proton. The lower blob represents the gluon distribution function in the proton, where two gluons are emitted with momentum fraction $x$ and transverse momentum $\boldsymbol{k}$.}
\end{figure}

The Fig.~\ref{fig:foto-part} shows the Feynman diagram for the $\gamma^{*}q$ subprocess, which represents only one contribution for the process, other possibilities are obtained exchanging the fermion lines in the color dipole. Moreover, the central exclusive production by the gluon-annihilation vertex yields four distinct diagrams, all of them needed to fully account for the process. In Ref. \cite{bialas} it is shown that the sum of the four possible diagrams results in the discontinuity of the diagram shown in Fig.~\ref{fig:foto-part}.
The two-upper blobs represent the effective vertices of the photon-gluon coupling which can be obtained through the impact factor formalism. The same formalism is used to explore the process with a non-zero momentum transfer with two gluons exchanged in the $t$-channel \cite{forshaw-evanson}. The other blob represents the gluon distribution function in the proton.

\begin{figure}[ht!]
\includegraphics[bb = 72 181 494 455 , scale=00.60]{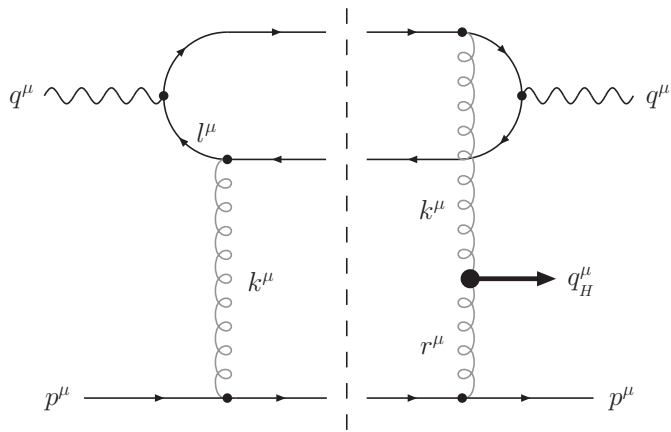}
\caption{\label{fig:foto-h} Diagram representing the diffractive Higgs boson photoproduction. The quarks circulating into the dipole have momenta $l^{\mu}$ and $q^{\mu}-l^{\mu}$. The photon impact factor is calculated in the $t = 0$ limit, and the coupling of the gluons ($k^{\mu},r^{\mu}$) to the proton is treated in the eikonal approximation.}
\end{figure}

The process that we study in this paper is based on the partonic subprocess $\gamma^{*} q \to \gamma^{*} + H + q$ shown in Fig.~\ref{fig:foto-h}. The central line cuts the diagram and expresses the use of the Cutkosky rules in order to obtain the imaginary part of the scattering amplitude, which is given by
\begin{eqnarray}
{\rm{Im}}A = \frac{1}{2} \int d(PS)_{3} \, {\cal{A}}_{L} \, {\cal{A}}_{R}
\label{cut-rule}
\end{eqnarray}
with ${\cal{A}}_{L}$ and ${\cal{A}}_{R}$ being the amplitudes on the left- and right-hand side of the cut, respectively, and $d(PS)_{3}$ is the volume element of the three-body phase space. The scattering amplitude of the process is treated essentially as an imaginary quantity in view of the vacuum quantum numbers of the exchanged particle \cite{foldy}. Moreover, the fermion loop is divided in two distinct parts, each other representing the splitting of the photon. In the Dipole Model \cite{mueller1} the splitting of the photon into a quark-antiquark pair requires an wave function, where its product with the complex conjugate represents the fermion loop. In the photoproduction approach, the impact factor formalism is used to describe the color dipole.

The representation of the partonic subprocess by the diagram in Fig.~\ref{fig:foto-h} shows a single process of $\gamma^{*}q$ interaction. The other diagrams can be obtained exchanging the coupling of the gluon lines to each fermion of the loop, and so the effective vertices are calculated with the help of the Feynman rules
\begin{subequations}
\begin{eqnarray}
\chi^{\mu\lambda} &=& \chi^{\mu\lambda}_{(L)} + \chi^{\mu\lambda}_{(R)} = i g_{s} e e_{q} (t^{a})_{{\cal{A}}{\cal{B}}} \left\{ (\gamma^{\mu})_{ij} \left[ \frac{(\llslash - \qslash)_{jk}}{(l_{1} - q)^{2}} \right] (\gamma^{\lambda})_{kl} + (\gamma^{\lambda})_{lk} \left[ \frac{(\llslash - \kslash)_{kj}}{(l_{1} - k)^{2}} \right] (\gamma^{\mu})_{ji} \right\},
\end{eqnarray}
and
\begin{eqnarray}
\chi^{\nu\sigma} &=& \chi^{\nu\sigma}_{(L)} + \chi^{\nu\sigma}_{(R)} = i g_{s} e e_{q} (t^{b})_{{\cal{B}}{\cal{A}}} \left\{ (\gamma^{\sigma})_{mn} \left[ \frac{(\kslash - \lllslash)_{np}}{(k - l_{2})^{2}} \right] (\gamma^{\nu})_{pq} + (\gamma^{\nu})_{qp} \left[ \frac{(\qslash - \lllslash)_{pn}}{(q - l_{2})^{2}} \right] (\gamma^{\sigma})_{nm} \right\},
\end{eqnarray}
\label{eff-vertex}
\end{subequations}
\hspace*{-.15cm}where the indices obey the assignment as follows: ($\mu, \nu, \sigma, ...$) for the four-vectors, ($a, b$) are the color indices, ($i, j, k, ...$) are the matrix-elements of the four-vectors, and (${\cal{A}}, {\cal{B}}, {\cal{C}}, ...$) for the elements of the color matrices.

\begin{figure}[ht!]
\includegraphics[bb = 27 293 485 474 , scale=00.70]{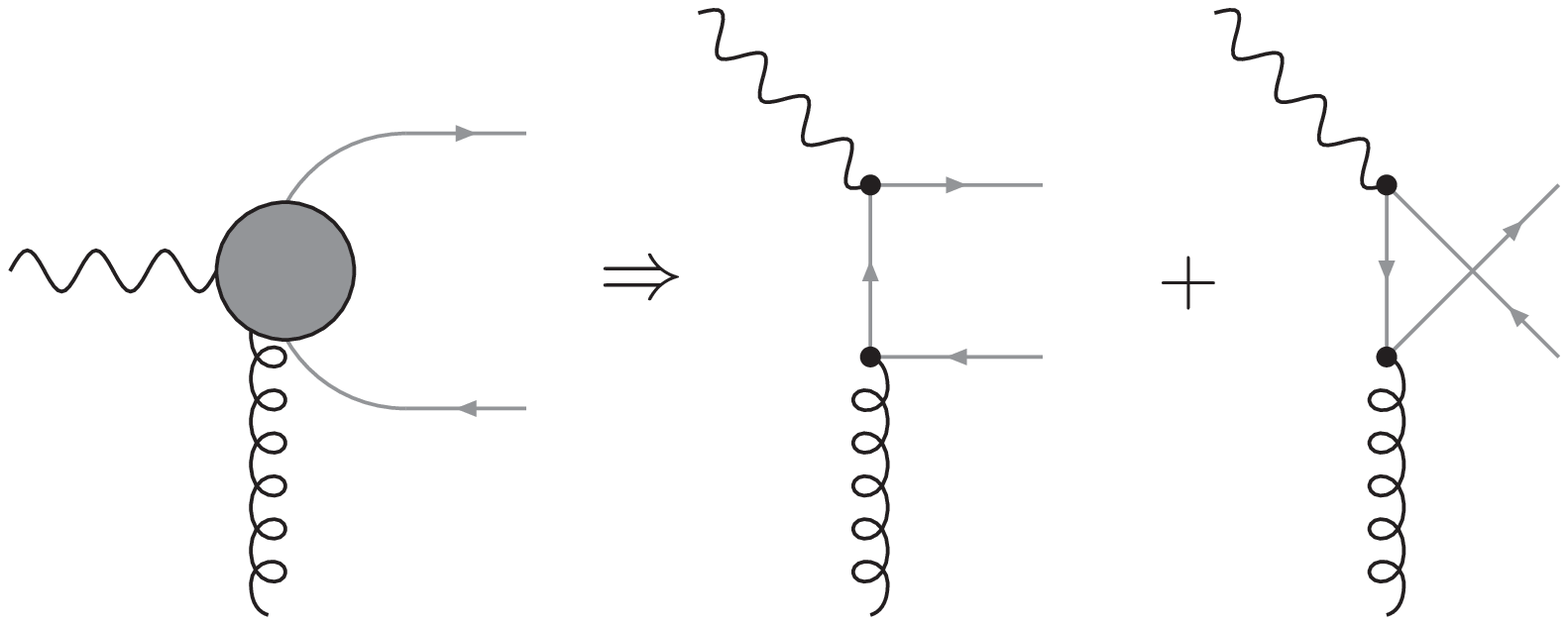}
\caption{\label{fig:eff-diag} Diagrams that contribute to the effective photon-gluon vertex. The product of its amplitude with the complex conjugate results in a diagram for the fermion loop. The sum of the four possibilities results in the whole contribution for the color dipole.}
\end{figure}

Attaching each side of the diagram, there will be two distinct contributions to the fermion loop, which can be computed whether one couples the diagrams in the Fig.~\ref{fig:eff-diag} at each side of the central line. The product of the diagrams in the left-hand side with the right- ones results in a possible diagram for the fermion loop, being the latter the complex conjugate of the former. This physical process is similar to that obtained in the Dipole Model: the wave function describes the photon splitting and its subsequent sprouting. Preventing an unnecessary calculation, one needs only to take into account two of the diagrams, since the other ones lead to the same contributions.

Computing the imaginary part of the amplitude defined in Eq.(\ref{cut-rule}), the product of the amplitudes in the left- and right-hand side is given by
\begin{eqnarray} \nonumber
{\cal{A}}_{L} {\cal{A}}_{R} &=& (4 \pi)^{3} \, \alpha^{2}_{s} \, \alpha \, \sum_{q} e^{2}_{q} \, \left( \frac{\epsilon_{\mu}\epsilon^{*}_{\nu}}{k^{4}r^{2}} \right) \frac{V^{ba}_{\delta\sigma}}{N_{c}} (t^{b}t^{a}) \, 4 p_{\lambda}p^{\sigma} \\
&\times& 2 \, \left\{ \frac{\textrm{Tr} \left[ (\qslash - \lslash) \gamma^{\mu} \lslash \gamma^{\lambda} (\kslash + \lslash) \gamma^{\sigma} \lslash \gamma^{\nu} \right]}{l^{4}} + \frac{\textrm{Tr} \left[ (\qslash - \lslash) \gamma^{\lambda} (\kslash + \lslash - \qslash) \gamma^{\mu} (\kslash + \lslash) \gamma^{\sigma} \lslash \gamma^{\nu} \right]}{l^{2}(k + l + q)^{2}} \right\},
\label{amp-cut-fhiggs}
\end{eqnarray}
where $\epsilon_{\mu}$ and $\epsilon^{*}_{\nu}$ are the polarization vectors of the initial and final photons, respectively, the vector $l^{\mu}$ is the four-momentum of the quark circulating into the fermion loop, and $p^{\mu}$ is the four-momentum of the colliding proton. Mathematically, the traces represent the fermion loop, which can be calculated with a numerical algorithm \cite{fORM}. The Gell-Mann $t$-matrices will appear as a trace of the color matrices when the product of ${\cal{A}}_{L}{\cal{A}}_{R}$ with its complex conjugate is performed. The quantity $V^{ba}_{\delta\sigma}$ represents the $ggH$ vertex, which is known as \cite{kniehl}
\begin{eqnarray}
V^{ab}_{\mu \nu} = \delta^{ab} \left( g_{\mu \nu} - \frac{k_{2\mu} k_{1\nu}}{k_{1} \cdot k_{2}} \right) V, \qquad\qquad\qquad V = F \left( \frac{M^{2}_{H}}{m^{2}_{t}} \right) \; \frac{M^{2}_{H} \alpha_{s}}{4 \pi v} \approx \frac{2}{3} \frac{M^{2}_{H} \alpha_{s}}{4 \pi v}.
\end{eqnarray}
The approximation for $F(x)$ is valid for the production of a non-heavy Higgs boson ($M_{H} \lesssim 200\textrm{ GeV}$).

However, the value of the traces involving a product of Dirac $\gamma$-matrices is obtained adopting a particular parametrization to the four-momenta presented in this process. In this way, the Sudakov parametrization is adopted, where the four-momenta are decomposed under three base-vectors: two light-type vectors $p^{\mu}$ and $q^{\prime \mu}$, where $q^{\prime \mu} = q^{\mu} + x p^{\mu}$, and a third vector lying in the plane perpendicular to the incident axis. The main kinematical variables are
\begin{eqnarray}
s = (q + p)^{2} \hspace*{2.cm} \hat{x} = Q^{2} / \, 2 \, (p \cdot q) \approx Q^{2}/s,
\end{eqnarray}
where $s$ is the squared center-of-mass energy of the photon-quark system, $\hat{x}$ is the Bjorken variable and $Q^{2} = - q^{2}$ is the photon virtuality. Thus, the decomposed four-momenta can be written as
\begin{subequations}
\begin{eqnarray}
\ell^{\mu} &=& \alpha_{\ell} q^{\prime \mu} + \beta_{\ell} p^{\mu} + \ell^{\mu}_{\perp} \\
k^{\mu} &=& \alpha_{k}q^{\prime \mu} + \beta_{k} p^{\mu} + k^{\mu}_{\perp} \\
r^{\mu} &=& \alpha_{r} q^{\prime \mu} + \beta_{r} p^{\mu} + r^{\mu}_{\perp}.
\end{eqnarray}
\end{subequations}
This set of decomposed four-vectors enables one to rewrite the denominators under the traces in Eq.(\ref{amp-cut-fhiggs}) as
\begin{subequations}
\begin{eqnarray}
\l^{2} = - \left[ \frac{\alpha_{\ell}(1 - \alpha_{\ell})Q^{2} + \boldsymbol{l}^{2}}{1 - \alpha_{\ell}} \right] &\equiv& - \frac{D_{1}}{1 - \alpha_{\ell}} \\
(l + k - q)^{2} = - \left[ \frac{\alpha_{\ell}(1 - \alpha_{\ell})Q^{2} + (\boldsymbol{l} + \boldsymbol{k})^{2}}{\alpha_{\ell}} \right] &\equiv& - \frac{D_{2}}{\alpha_{\ell}}.
\end{eqnarray}
\label{d1d2}
\end{subequations}

The final step is to write out the volume element of the three-body phase space under the Sudakov parametrization in order to obtain the imaginary part of the scattering amplitude. The definition of the volume element reads
\begin{eqnarray} \nonumber
\int d(PS)_{3} &=& \int \frac{d^{4}f_{1}}{(2\pi)^{3}} \, \frac{d^{4}f_{2}}{(2\pi)^{3}} \, \frac{d^{4}f_{3}}{(2\pi)^{3}} \, \delta(f^{2}_{1}) \, \delta(f^{2}_{2}) \, \delta(f^{2}_{3}) \, (2 \pi)^{4} \, \delta^{4}(q + p - f_{1} - f_{2} - f_{3}) \\
&=& \frac{1}{(2 \pi)^{5}} \int d^{4}l \, d^{4}k \, \delta([q - \ell]^{2}) \, \delta([\ell + k]^{2}) \, \delta([p - k]^{2}),
\label{dps}
\end{eqnarray}
which expressed under the Sudakov parametrization reads
\begin{eqnarray}
\int d(PS)_{3} = \int d\alpha_{\ell} \, d\beta_{\ell} \, d^{2}\boldsymbol{l} \int d\alpha_{k} \, d\beta_{k} \, d^{2}\boldsymbol{k} \;\; \delta \left[ \beta_{\ell} + \frac{Q^{2}}{s} + \frac{\boldsymbol{l}^{2}}{s(1 - \alpha_{\ell})} \right] \delta \left[ \beta_{k} + \frac{(\boldsymbol{l} + \boldsymbol{k})^{2}}{\alpha_{\ell} s} + \beta_{\ell} \right] \, \delta [\alpha_{k} s + \boldsymbol{k}^{2}].
\end{eqnarray}

Considering the Cutkosky rules to calculate the scattering amplitude, the quarks are on-mass shell due to the delta functions in Eq.(\ref{dps}), which allows one to perform the following approximation to the gluon momentum
\begin{eqnarray}
k^{2} \simeq - \boldsymbol{k}^{2} \hspace{2.cm} r^{2} \simeq - \boldsymbol{r}^{2} \approx - \boldsymbol{k}^{2}.
\end{eqnarray}
Physically, this approximantion is an important feature, since it accesses the kinematical region of the $H \to b\bar{b}$ decay mode and experimentally expresses that the quarks are scattered in small angles \cite{forshaw-KMR}.

The integration of the delta functions results in the following imaginary part of the amplitude
\begin{eqnarray}
\textrm{Im} A = V \left( \frac{2\alpha_{s}^{2} \, \alpha}{\pi^2 s} \right) \sum_{q} e^{2}_{q} \left( \frac{\epsilon_{\mu} \epsilon^{*}_{\nu}}{N_{c}} \right) (t^{a}t^{a}) \int d\alpha_{\ell} \, \frac{d^{2}\boldsymbol{k}}{\boldsymbol{k}^{6}} \, d^{2}\boldsymbol{l} \left[ \frac{(1 - \alpha_{\ell})}{\alpha_{\ell}} \frac{T^{\mu\lambda\sigma\nu}}{(D_{1})^{2}} + \frac{T^{\lambda\mu\sigma\nu}}{D_{1}D_{2}} \right] \left[ p_{\lambda}p_{\sigma} - \frac{(k \cdot p)}{\boldsymbol{k}^{2}} p_{\lambda} r_{\sigma} \right],
\label{amp-img-fhiggs}
\end{eqnarray}
with the quantities $T^{\mu\lambda\sigma\nu}$ and $T^{\lambda\mu\sigma\nu}$ being the traces present in Eq.(\ref{amp-cut-fhiggs}), and the decomposed vectors acquire its coefficients from the integration of the delta functions
\begin{subequations}
\begin{eqnarray}
l^{\mu} &=& \alpha_{\ell} q^{\prime \mu} - \left( Q^{2} + \frac{\boldsymbol{l}^{2}}{1 - \alpha_{\ell}} \right) \frac{p^{\mu}}{s} + l^{\mu}_{\perp} \\
k^{\mu} &=& - \frac{\boldsymbol{k}^{2}}{s} q^{\prime \mu} + \left[ Q^{2} + \frac{\boldsymbol{l}^{2}}{1 - \alpha_{\ell}} + \frac{(\boldsymbol{l} + \boldsymbol{k})^{2}}{\alpha_{\ell}} \right] \frac{p^{\mu}}{s} + k^{\mu}_{\perp}.
\end{eqnarray}
\end{subequations}
For completeness, one can write the momentum transfer of the process as
\begin{eqnarray}
\hat{t} = (k + r)^{2} \equiv - (\boldsymbol{k} + \boldsymbol{r})^{2} \approx - 4 \boldsymbol{k}^{2} = M^{2}_{H}.
\end{eqnarray}

In this proposal of Higgs boson photoproduction, the last variables to be determined are the polarization vectors of the initial and final photons. As said in the Sec. I, we apply the impact factor of the photon for $t = 0$. Thus, the polarization vectors do not depend on the $t$-variable, being its sum over transversal\footnote{The transversal mode and the transversal vectors lie in the plane perpendicular to the incident axis, however they have different meanings. To avoid confusion, the scattering amplitude in transversal mode implies that the transversal polarization mode was assumed in the calculation. There is no mention to the transversal vectors, which are integrated at the end.} and longitudinal components expressed by \cite{forshaw-book}
\begin{subequations}
\begin{eqnarray}
\epsilon^{L}_{\mu} \, \epsilon^{L *}_{\nu} &=& \frac{4Q^{2}}{s} \, \frac{p_{\mu}p_{\nu}}{s} \\
\sum \epsilon^{T}_{\mu} \, \epsilon^{T *}_{\nu} &=& - g_{\mu\nu} + \frac{4Q^{2}}{s} \, \frac{p_{\mu}p_{\nu}}{s}.
\end{eqnarray}
\end{subequations}
These relations complete the set of information necessary to fully compute the imaginary part of the scattering amplitude for transversal and longitudinal modes. Nevertheless, in this kinematical regime, $\alpha$ is approximately equal to 1, and $\alpha_{k}$ can be neglected as well as the product $(k \cdot p)$ present in the Eq.(\ref{amp-img-fhiggs}) \cite{kuchina}.

Hence, the two modes of the imaginary part of the scattering amplitude are given by
\begin{subequations}
\begin{eqnarray}
(\textrm{Im}A)_{L} \simeq \left( \frac{1}{\pi s^{3}} \right) 4Q^{2} \, \alpha_{s}^{2} \, \alpha \sum_{q} e^{2}_{q} \left[ \frac{N^{2}_{c} - 1}{N^{2}_{c}} \right] V \int d\alpha_{\ell} \; d^{2}\boldsymbol{l} \; \frac{d^{2}\boldsymbol{k}}{\boldsymbol{k}^{6}} \left[ \frac{(1 - \alpha_{\ell})\Gamma_{1}}{\alpha_{\ell}(D^{2}_{1})} + \frac{\Gamma_{2}}{D_{1}D_{2}} \right],
\end{eqnarray}and
\begin{eqnarray}
\!\!\!\!\!\!\!\! (\textrm{Im}A)_{T} \simeq \frac{V}{\pi^{2} s^{3}} \sum_{q} e^{2}_{q} \left[ \frac{N^{2}_{c} - 1}{N^{2}_{c}} \right] \!\! \int d\alpha_{\ell} \; d^{2}\boldsymbol{l} \; \frac{d^{2}\boldsymbol{k}}{\boldsymbol{k}^{6}} \!\! \left[ \frac{(1 - \alpha_{\ell})(-s^{2} g_{\mu\nu} \, \Gamma^{\mu\nu}_{1} + 4Q^{2}\pi \Gamma_{1} )}{\alpha_{\ell}D^{2}_{1}} + \frac{\left( - s^{2} g_{\mu\nu} \, \Gamma^{\mu\nu}_{2} + 4Q^{2}\pi \Gamma_{2} \right)}{D_{1}D_{2}} \right] ,
\end{eqnarray}
\end{subequations}
where $\Gamma_{i}$ and $\Gamma^{\mu\nu}_{i}$ are trace functions of the form
\begin{subequations}
\begin{eqnarray}
\Gamma_{1} &=& \textrm{Tr} \left[ (\hspace*{-00.10cm}\qslash - \hspace*{-00.20cm} \lslash) \hspace*{-00.10cm} \pslash \hspace*{-00.10cm} \lslash \hspace*{-00.10cm} \not{\hbox{\kern-2.5pt $p$}} \hspace*{00.05cm} (\hspace*{-00.10cm}\lslash + \hspace*{-00.20cm} \kslash) \hspace*{-00.10cm} \pslash \hspace*{-00.10cm} \lslash \hspace*{-00.05cm} \pslash \hspace*{00.05cm} \right] \\
\Gamma_{2} &=& \textrm{Tr} \left[ (\hspace*{-00.10cm}\qslash - \hspace*{-00.20cm} \lslash) \hspace*{-00.10cm} \pslash \hspace*{00.10cm} (\hspace*{-00.10cm}\kslash + \hspace*{-00.20cm} \lslash - \hspace*{-00.20cm} \qslash) \hspace*{-00.10cm} \pslash \hspace*{00.10cm} (\hspace*{-00.10cm}\kslash + \hspace*{-00.20cm} \lslash) \hspace*{-00.10cm} \pslash \hspace*{-00.10cm} \lslash \hspace*{-00.05cm} \not{\hbox{\kern-2.5pt $p$}} \hspace*{00.05cm} \right] \\
\Gamma^{\mu\nu}_{1} &=& \textrm{Tr} \left[ (\hspace*{-00.10cm}\qslash - \hspace*{-00.20cm} \lslash) \hspace*{00.05cm} \gamma^{\mu} \hspace*{-00.15cm} \lslash \hspace*{-00.05cm} \pslash \hspace*{00.10cm} (\hspace*{-00.10cm} \lslash + \hspace*{-00.20cm} \kslash) \hspace*{-00.10cm} \pslash \hspace*{-00.10cm} \lslash \hspace*{00.10cm} \gamma^{\nu} \right] \\
\Gamma^{\mu\nu}_{2} &=& \textrm{Tr} \left[ (\hspace*{-00.10cm}\qslash - \hspace*{-00.20cm} \lslash) \hspace*{-00.10cm} \pslash \hspace*{00.10cm} (\hspace*{-00.10cm} \kslash + \hspace*{-00.20cm} \lslash - \hspace*{-00.20cm} \qslash) \hspace*{00.05cm} \gamma^{\mu} \hspace*{00.05cm} (\hspace*{-00.10cm} \kslash + \hspace*{-00.20cm} \lslash) \hspace*{-00.11cm} \pslash \hspace*{-00.10cm} \lslash \hspace*{00.10cm} \gamma^{\nu} \right].
\end{eqnarray}
\end{subequations}
Computing these traces, the transversal mode of the scattering amplitude results
\begin{eqnarray}
(\textrm{Im}A)_{T} = \left( \frac{V}{\pi^{2}} \right) \alpha_{s}^{2} \, \alpha \sum_{q} e^{2}_{q} \left( \frac{2C_{F}}{N_{c}} \right) \int^{1}_{0} d\alpha_{\ell} \int^{\infty}_{0} d^{2}\boldsymbol{l} \; \frac{d^{2}\boldsymbol{k}}{\boldsymbol{k}^{6}} \left[ \frac{\xi_{1}}{D^{2}_{1}} + \frac{\xi_{2}}{D_{1}D_{2}} \right] \hspace*{-.1cm},
\end{eqnarray}
with $D_{1}$ and $D_{2}$ defined in Eq.(\ref{d1d2}), and the terms $\xi_{1}$ and $\xi_{2}$ read
\begin{subequations}
\begin{eqnarray}
\xi_{1} &=& 4Q^{2} \alpha_{\ell} (1 - \alpha_{\ell}) (1 - \alpha_{\ell} + \alpha^{2}_{\ell})s,
\end{eqnarray}and
\begin{eqnarray}
\xi_{2} &=& - \left[ 4\boldsymbol{k}^{2} + 4Q^{2}\alpha_{\ell} (1 - \alpha_{\ell}) \right] (1 - \alpha_{\ell} + \alpha_{\ell}^{2})s - 4(\boldsymbol{k} \cdot \boldsymbol{l})s = \xi^{\prime}_{2} - 4(\boldsymbol{k} \cdot \boldsymbol{l})s.
\end{eqnarray}
\end{subequations}

To perform the integration over the transversal vector $\boldsymbol{l}$, the Feynman parameter is introduced
\begin{eqnarray}
\frac{1}{AB} = \int^{1}_{0} \; \frac{1}{\left[ A + (B - A)\tau \right]^{2}} \, d\tau,
\end{eqnarray}
and one obtains the following results
\begin{eqnarray}
\int \frac{d^{2}\boldsymbol{l}}{D_{1}D_{2}} = \int^{1}_{0} \frac{d\tau}{\boldsymbol{k}^{2} (\tau - \tau^{2}) + Q^{2} \alpha_{\ell}(1 - \alpha_{\ell})},
\end{eqnarray}
and also the second integration can be performed
\begin{eqnarray}
\int d^{2} \boldsymbol{l} \; \frac{1}{[ \boldsymbol{l}^{2} + Q^{2}\alpha_{\ell}(1 - \alpha_{\ell}) ]^{2}} = \frac{\pi}{Q^{2}\alpha_{\ell}(1 - \alpha_{\ell})}.
\end{eqnarray}

After performing these integrations, one gets the scattering amplitude in transversal mode
\begin{eqnarray}
(\textrm{Im}A)_{T} = - \frac{s}{3} \left( \frac{M^{2}_{H}}{\pi v}  \right) \alpha_{s}^{3} \, \alpha \sum_{q} e^{2}_{q} \left( \frac{2C_{F}}{N_{c}} \right) \int \frac{d\boldsymbol{k}^{2}}{\boldsymbol{k}^{6}} \left\{ \int_{0}^{1} \frac{[\tau^{2} + (1 - \tau)^{2}][\alpha_{\ell}^{2} + (1 - \alpha_{\ell})^{2}]\boldsymbol{k}^{2}}{\boldsymbol{k}^{2} \tau (1 - \tau) + Q^{2} \alpha_{\ell}(1 - \alpha_{\ell})} \; d\alpha_{\ell} d\tau \right\}
\label{amp-im-1}
\end{eqnarray}
being $v = 246\textrm{ GeV}$ the vacuum expectation value (v.e.v.) from the Higgs mechanism. Following the same procedure, the longitudinal mode of the amplitude reads
\begin{eqnarray}
(\textrm{Im}A)_{L} = \frac{4s}{3} \left( \frac{M^{2}_{H}}{\pi v}  \right) \alpha_{s}^{3} \, \alpha \sum_{q} e^{2}_{q} \left( \frac{2C_{F}}{N_{c}} \right) \int \frac{d\boldsymbol{k}^{2}}{\boldsymbol{k}^{6}} \left\{ \int_{0}^{1} \frac{[\tau(1 - \tau)][\alpha_{\ell}(1 - \alpha_{\ell})]\boldsymbol{k}^{2}}{\boldsymbol{k}^{2} \tau (1 - \tau) + Q^{2} \alpha_{\ell}(1 - \alpha_{\ell})} \; d\alpha_{\ell} d\tau \right\}.
\label{amp-im-2}
\end{eqnarray}
Nonetheless, a longitudinal mode for a real photon is an unphysical property and then only the transversal one is taken into account. Integrating the Eq.(\ref{amp-im-1}) over $\alpha_{\ell}$ and $\tau$, one finds
\begin{eqnarray}
(\textrm{Im}A)_{T} = - \frac{4s}{9} \left( \frac{M^{2}_{H}}{\pi v}  \right) \alpha_{s}^{3} \, \alpha \sum_{q} e^{2}_{q} \left( \frac{2C_{F}}{N_{c}} \right) \int \frac{d\boldsymbol{k}^{2}}{\boldsymbol{k}^{6}} \left( 1 + \frac{24\boldsymbol{k}^{8}-226Q^{2}\boldsymbol{k}^{6}-733Q^{4}\boldsymbol{k}^{4}-670Q^{6}\boldsymbol{k}^{2}-186Q^{8}}{24Q^{8}+72Q^{6}\boldsymbol{k}^{2}+72Q^{4}\boldsymbol{k}^{4}+24Q^{2}\boldsymbol{k}^{6}} \right).
\label{amp-im-3}
\end{eqnarray}

Finally, only the transversal mode is retained to compute the event rate, which is expressed as a central-rapidity distribution of the Higgs boson ($y_{H} = 0$) through the relation $d^{3}\vec{q}_{H} = \pi E_{H} d\boldsymbol{q}^{2}_{H} dy_{H}$, then
\begin{eqnarray}
\left. \frac{d\sigma}{dy_{H}d\boldsymbol{p}^{2}dt} \right|_{t,y_{H} = 0} = \frac{1}{162 \pi^{4}} \left( \frac{M^{2}_{H}}{N_{c}v} \right)^{2} \alpha_{s}^{4} \, \alpha^{2} \left( \sum_{q} e^{2}_{q} \right)^{2} \left[ \frac{\alpha_{s} \, C_{F}}{\pi} \int \frac{d\boldsymbol{k}^{2}}{\boldsymbol{k}^{6}} \; {\cal{X}}(\boldsymbol{k}^{2},Q^{2}) \right]^{2},
\end{eqnarray}
where the function ${\cal{X}}(\boldsymbol{k}^{2},Q^{2})$ is the function inner the parenthesis in Eq.(\ref{amp-im-3}).

The main feature obtained in this result is the sixth-order $\boldsymbol{k}$-dependence, since it is distinct of the result of the Durham group, which presents a fourth-order dependence. Such difference happens due to the presence of the photon in the process, turning the result more simplified, although introducing a more complicated expression with a $Q^{2}$-dependence.

\section{PHOTON-PROTON COLLISIONS}\label{sec:proton}

\begin{figure}[ht!]
\includegraphics[bb = 26 529 203 778 , scale=00.55]{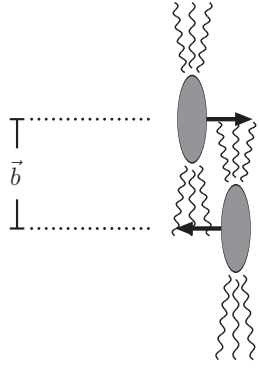}
\caption{\label{fig:periph} Presentation of the general dynamics of Peripheral Collisions: the wave lines represent the photon field of the hadrons under the Lorentz contraction, where the quantity $\vec{b}$ is the impact parameter of the process, which is considered to be $|\vec{b}| \gtrsim 2R$.}
\end{figure}

The main interest of this work is the Higgs boson production in a subprocess of Peripheral Collision \cite{bertu,hencken}, where a photon from one of the protons (ions) under collision can interact with the second one, which is shown in Fig.~\ref{fig:periph}. In this case, strong interactions do not occur due to the large distance between the partonic content of the hadrons, i.e., only the electromagnetic interaction can occur, being the basic assumption to consider the impact parameter to be $ |\vec{b}| > R_{1} + R_{2} \gtrsim 2R$. Often, these particles are called quasi-real or equivalent photons, and have an energy spectrum hardly dependent on collision energy \cite{bauretal}. Other important aspect is the dependence of the Coulomb field on the number of charged particles into the hadron, and considering nucleus-nucleus collisions, the equivalent photon number depends on
\begin{eqnarray}
n(\omega) \propto Z^{2},
\end{eqnarray}
yielding an important contribution in photon interactions.

For the Higgs photoproduction, the virtuality of the photon exchanged in the peripheral proton-proton collision needs to be estimated. The source of photons in this kind of process is the Coulomb field around the protons under collision, and the photon spectrum is calculated by the Equivalent Photon Method (EPA), obtaining the Weizs\"acker-Williams distribution \cite{bauretal,bertu}. The photons are soft particles that have low virtuality with an upper limit determined by the hadron radius
\begin{eqnarray}
Q^{2} \lesssim 1/R_{p}^{2},
\end{eqnarray}
which is $Q^{2} \lesssim$ 10$^{-2}$ GeV$^{2}$ in $pp$ collisions.

The calculations were performed at partonic level, i.e., considering the photon-quark interaction. A realistic case of photon-proton interaction in Peripheral Collisions is built if one replaces the contribution of the quark-gluon vertices by a partonic distribution in the proton, as shown by the lower blob of Fig.~\ref{fig:foto-part}. This distribution is considered as a non-integrated distribution function regarding the coupling of the gluons to the proton through a gluon ladder described by the BFKL equation \cite{KMR-CanThe}. Moreover, the diffractive Higgs production is assumed to be an elastic process, where the momenta of the gluons are much smaller than the other kinematical variables under consideration. For a realistic description of the process, the partonic distribution is assumed to be a non-diagonal distribution, to express a more general situation, such that the proton loses a small fraction of its momentum during the collision. The matrix considered for the parton distribution in the proton is not diagonal, which means that the proton vertex has non-zero momentum transfer \cite{off-diag}. Thus, the following replacement is performed to describe the $\gamma p$ interaction
\begin{eqnarray}
\frac{\alpha_{s} \, C_{F}}{\pi} \;\; \longrightarrow \;\; f_{g}(x,\boldsymbol{k}^{2}) = {\cal{K}} \left( \frac{\partial [xg(x,\boldsymbol{k}^{2})]}{\partial \ell n \, \boldsymbol{k}^{2}} \right)
\end{eqnarray}
where $f_{g}(x,\boldsymbol{k}^{2})$ is the non-diagonal gluon distribution function in the proton evolved by the BFKL equation. The non-diagonality of the distribution can be approximated by a multiplicative factor ${\cal{K}}$, which possesses a Gaussian shape \cite{shuvaevetal}
\begin{eqnarray}
{\cal{K}} = (1.2)\, \textrm{exp}(-B\boldsymbol{p}^{2}/2),
\end{eqnarray}
with $B$ being the impact parameter, assumed to be $B = 5.5\mbox{ GeV}^{-2}$ \cite{miller}. This factor can be seen as the representation of the proton-Pomeron coupling. However, in order to assume the gluon ladder coupled to the proton, the consideration of zero momentum transfer in the proton vertex is not a sufficient condition. A small value for the momentum fraction is required in this region of interest, like $x = M_{H}/\sqrt{s} \sim 0.01$, such that one can safely put $t_{p} = 0$, and identify the distribution as the unintegrated gluon distribution function $f_{g}(x,\boldsymbol{k}^{2})$ evolved by the BFKL evolution equation \cite{KMR-1997}.

Finally, the event rate has the form
\begin{eqnarray}
\left. \frac{d\sigma}{dy_{H}d\boldsymbol{p}^{2}dt} \right|_{t,y_{H} = 0} = \frac{(1.2)^{2} }{162 \pi^{4} } \left( \frac{M^{2}_{H}}{N_{c}v} \right)^{2} \alpha_{s}^{4} \, \alpha^{2} \left( \sum_{q} e^{2}_{q} \right)^{2} e^{-B\boldsymbol{p}^{2}} \left[ \int  \frac{d\boldsymbol{k}^{2}}{\boldsymbol{k}^{6}} \; f_{g}(x,\boldsymbol{k}^{2}) \; {\cal{X}}(\boldsymbol{k}^{2},Q^{2}) \right]^{2}.
\label{dpH}
\end{eqnarray}
For momentum conservation, there is a relation between the transversal components of the Higgs and the proton momenta, being
\begin{eqnarray}
d\boldsymbol{p}^{2}_{H} \longrightarrow - \, d\boldsymbol{p}^{2}.
\label{dpH2}
\end{eqnarray}
In Eq.(\ref{dpH}), one can perform the last integration over the transversal component of the proton momentum, resulting in the final expression for the diffractive production in $\gamma p$ interaction
\begin{eqnarray}
\left. \frac{d\sigma}{dy_{H}dt} \right|_{t,y_{H} = 0} = \frac{2 \alpha_{s}^{4} \, \alpha^{2} }{225 \pi^{4} \, b} \left( \frac{M^{2}_{H}}{N_{c}v} \right)^{2} \left( \sum_{q} e^{2}_{q} \right)^{2} \left[ \int \frac{d\boldsymbol{k}^{2}}{\boldsymbol{k}^{6}} \; f_{g}(x,\boldsymbol{k}^{2}) \; {\cal{X}}(\boldsymbol{k}^{2},Q^{2}) \right]^{2}.
\end{eqnarray}

An important feature considered by the Durham group is the suppression of the gluon emissions from the production vertex, i.e., gluons \textit{bremsstrahlung} \cite{KMR-Rates}. The suppression probability $S$ for the emission of one gluon can be computed with the help of Sudakov form factors, such that
\begin{eqnarray}
S(\boldsymbol{k}^{2},M^{2}_{H}) = \int^{M^{2}_{H}/4}_{\boldsymbol{k}^{2}} \frac{C_{A} \alpha_{s}(p^{2}_{T})}{\pi} \, \frac{dp^{2}_{T}}{p^{2}_{T}} \int^{M_{H}/2}_{p_{T}} \frac{dE}{E} = \frac{3 \alpha_{s}}{4 \pi} \ell n^{2} \left( \frac{M^{2}_{H}}{4\boldsymbol{k}^{2}} \right)
\label{sud-supress}
\end{eqnarray}
where $E$ and $p_{T}$ are the energy and the transversal momentum of an emitted gluon in the rest-frame of the Higgs boson, respectively. The above result is obtained using a fixed strong coupling constant in the integration. The suppression of many emissions exponentiates, and an exponential term is introduced to the event rate
\begin{eqnarray}
\left. \frac{d\sigma}{dy_{H}dt} \right|_{t,y_{_{H}} = 0} = \frac{1}{18 \pi^{3} b } \left( \frac{M^{2}_{H}}{N_{c}v} \right)^{2} \alpha_{s}^{4} \, \alpha^{2} \left( \sum_{q} e^{2}_{q} \right)^{2} \left[ \int_{\boldsymbol{k}^{2}_{0}}^{\infty} \frac{d\boldsymbol{k}^{2}}{\boldsymbol{k}^{6}} \; e^{-S(\boldsymbol{k}^{2},M^{2}_{H})} \; f_{g}(x,\boldsymbol{k}^{2}) \; {\cal{X}}(\boldsymbol{k}^{2},Q^{2}) \right]^{2},
\label{final-eq1}
\end{eqnarray}
where a cutoff was included in the integration on the gluon momentum to avoid infrared divergences \cite{forshaw-KMR}.

A last important aspect regarded to the diffractive process is the rapidity gaps present in the final state due to the vacuum quantum numbers of the exchanged particle: the Pomeron. It means that the final state has a particular rapidity distribution, where the rapidity range between the colliding particles is free of secondary particles, i.e., there is no production of particles in this region, only the Higgs. However, the rapidity gaps predicted theoretically are bigger than those observed in the experimental results. This contradiction occurs due to the still poor theoretical description of the interactions occuring by the presence of secondary particles. The mechanism that provides the correct prediction of the rapidity gaps is the Rapidity Gap Survival Probability (GSP), which accounts for the probability that, during a process, the rapidity gap will survive to interactions with the spectator particles. In other words, the GSP is the probability that will have an event where does not occur other interactions except the hard collision. Thus, a multiplicative factor $S^{2}_{gap}$ is included in Eq.(\ref{final-eq1}), which will account for the reduction of the predicted cross section, reaching the correct value expected to be measured in the laboratory.

The survival probability was originally formulated by Bjorken \cite{bjorken1993} as
\begin{eqnarray}
S^{2}_{gap} = \frac{\int d^{2}b \, \Gamma_{H} (b) \, | P(s,b) |^{2}}{\int d^{2}b \, \Gamma_{H}(b)},
\end{eqnarray}
where $\Gamma_{H}(b)$ is the profile function and $P(s,b)$ is the probability that inelastic interactions occur during the process. For this proposal, the assumptions of our paper are based in the previous works that calculate the GSP for Higgs production  \cite{chehime,KMR,luna,gotsman}. As a kinematical consequence, it has a dependence on the center-of-mass energy of the process, such that it decreases as the energy increases. Consequently, the Durham group estimates the survival probability for diffractive Higgs production considering a similar approach than that of Bjorken. Their approach consists in computing the GSP through
\begin{eqnarray}
S^{2}_{gap} = \frac{\int |{\cal{M}}(s,b)|^{2} \, e^{-\Omega(b)} \, d^{2}b}{\int |{\cal{M}}(s,b)|^{2} \, d^{2}b},
\end{eqnarray}
where ${\cal{M}}(s,b)$ is the scattering amplitude of the process in the impact-parameter space at the squared center-of-mass energy $s$. The function $\Omega(b)$ is the opacity (or optical density) of the interaction between the hadrons under collision. A relevant feature of this approach is that the GSP depends on the particular hard subprocess under study, and its kinematical configurations \cite{KMR}. Furthermore, there is a dependence of the GSP on the parton distribution in the protons, which is described in the impact-parameter space, and can be parametrized by several proposals \cite{GRV,MRST,ALEKHIN,CTEQ}. As a result of this approach, the Durham group estimates the survival probability for the diffractive Higgs production to be 3\% in LHC ($\sqrt{s} = 14\textrm{ TeV}$) and 5\% in Tevatron ($\sqrt{s} = 1.96\textrm{ TeV}$).

\section{NUMERICAL RESULTS}\label{sec:results}

The results for the Higgs photoproduction are obtained with the help of a set of parametrizations of the gluon distribution function in the proton \cite{lhapdf}.
\begin{figure}[ht!]
\centering
\resizebox{\textwidth}{!}{
\rotatebox{-90}{\scalebox{00.21}{\includegraphics*[86pt,25pt][568pt,725pt]{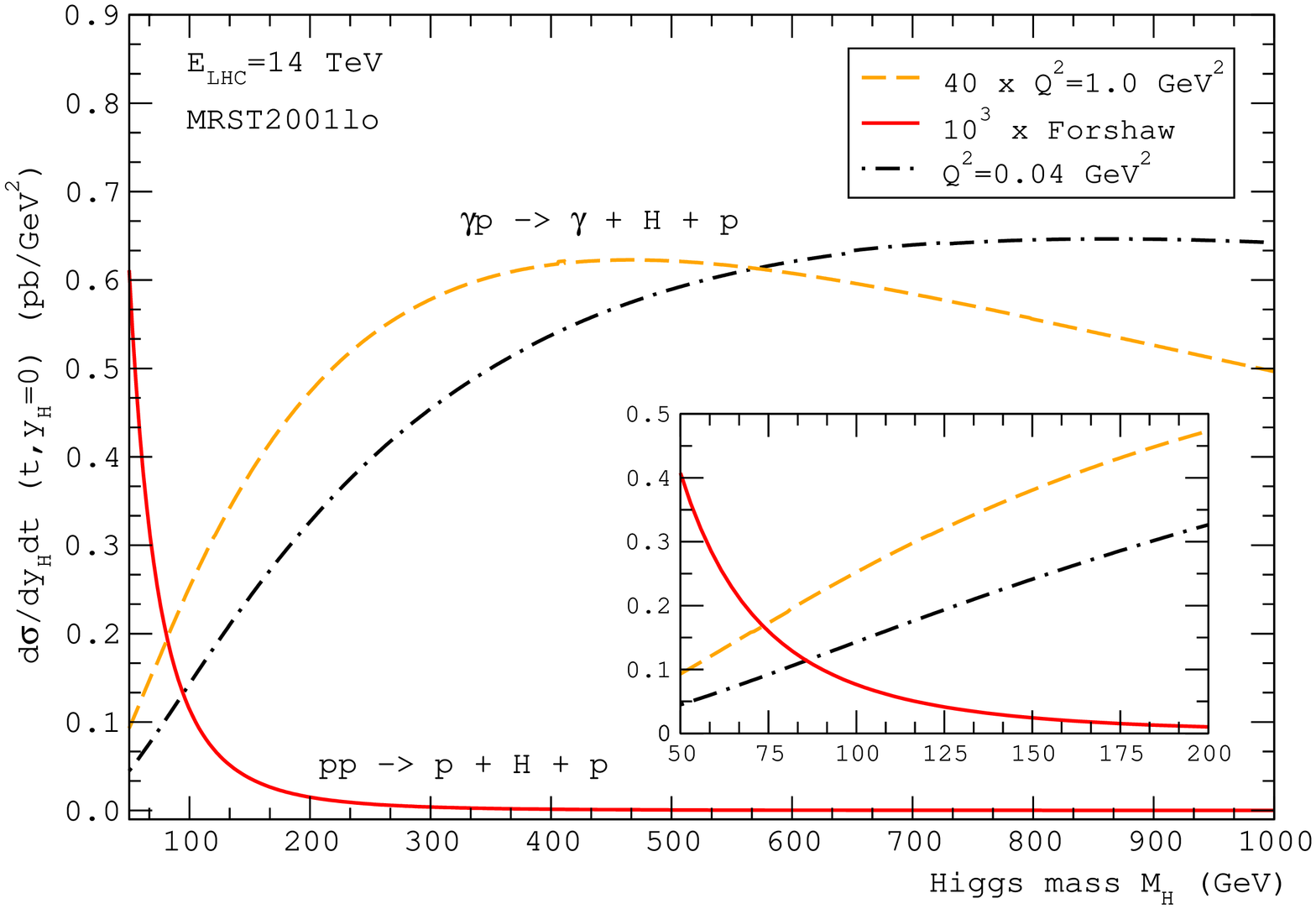}}}
\hfill
\rotatebox{-90}{\scalebox{00.20}{\includegraphics*[81pt,23pt][575pt,717pt]{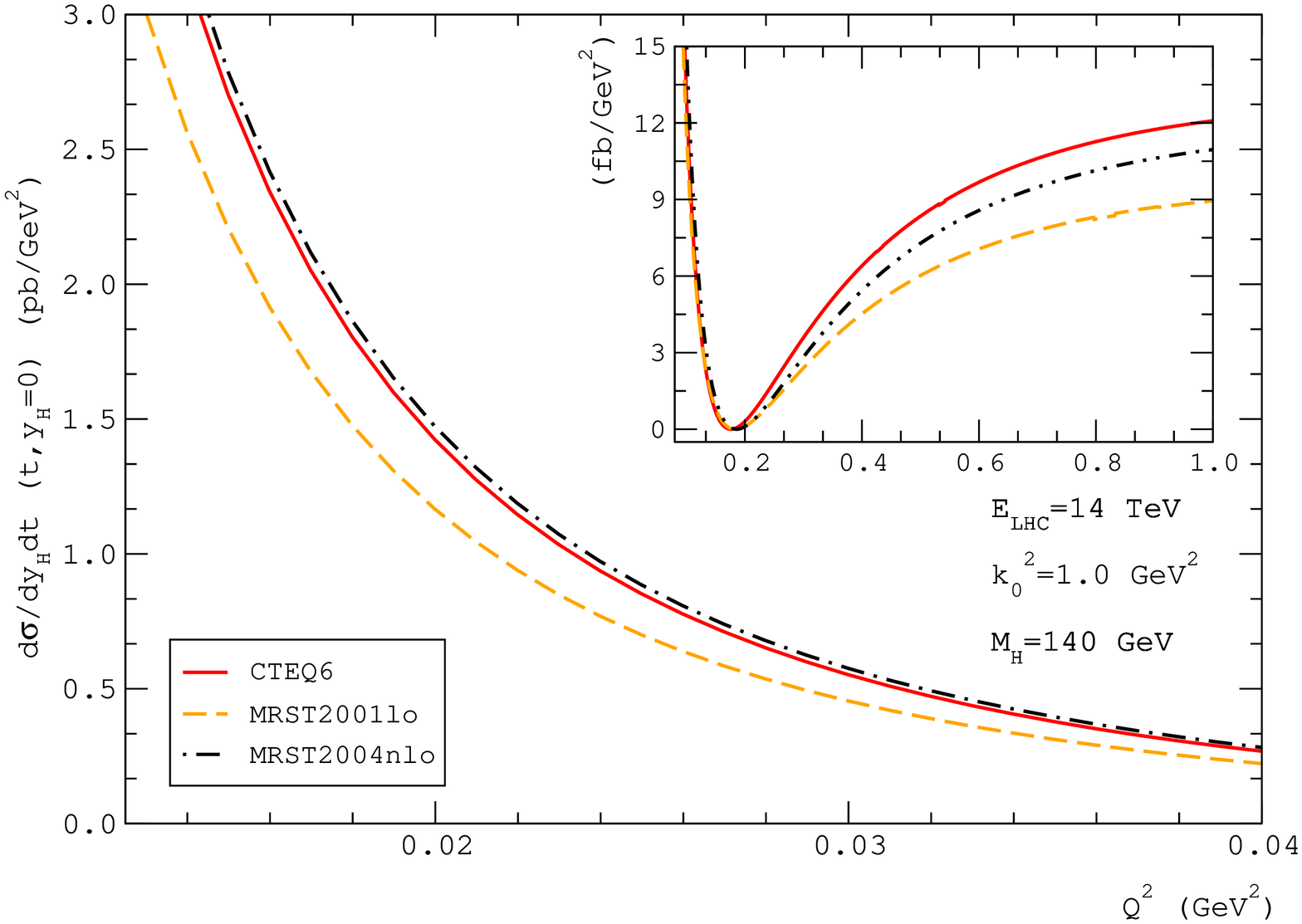}}}
}
\caption{\label{fig:forshaw} On the left-hand side: event rate $ d\sigma/dy_{H}$ ($y_{H}$=0) for LHC energy in function of the Higgs mass in two distinct ranges: the smaller figure shows the results in the intermediary mass range, and the bigger figure  presents the extended range. The results are obtained using the MRST2001 parametrization in LO approximation. The curves show the predictions for different virtualities: Q$^{2}$ = 1.0 GeV$^{2}$ (re-scaled) and Q$^{2}$=0.04 GeV$^{2}$. These results are compared with the previous prediction of the Durham group carried out by Forshaw (solid line). On the right-hand side: the graph shows the dependence of the event rate on the photon virtuality for different parametrizations.}
\end{figure}

The first step is to compare the results obtained for the photoproduction approach with those for the Higgs production in direct $pp$ collisions carried out in Ref.~\cite{forshaw-KMR}. Hence, the prediction for the differential cross section in central-rapidity for LHC is calculated using the parametrization MRST2001 in Leading Order (LO) approximation for the gluon distribution function, taking the initial momentum at $\boldsymbol{k}^{2}_{0} = 1.0 \textrm{ GeV}^{2}$. The results are expressed in Fig.~\ref{fig:forshaw}, where the event rate is fitted in function of the Higgs boson mass. The smallest figure represents the result in the mass range where the approximation to $F(x)$, present in the $gg \to H$ production vertex, is valid. As the mass increases, the curve seems to diverge, however extending the mass range to largest values of the Higgs mass, one sees that this growth is reduced, which is caused by the presence of the Sudakov form factors. A distinct behavior between the photoproduction results and those of direct $pp$ collision is present in virtue of the different approaches implemented in both cases. In the direct $pp$ collision approach, there are two distribution functions expressing the content of the two interacting protons, while the photoproduction possesses only one distribution function. Thus, the behavior of the photoproduction results is expected not to fit like the results of direct $pp$ collisions. Analysing the dependence on the photon virtuality, one sees a fast decreasing of the event rate in the range below Q$^{2} \approx$ 0.2 GeV$^{2}$. In the extended range of photon virtuality, the behavior of the event rate is fitted up to Q$^{2}$ = 1.0 GeV$^{2}$, showing a fast decreasing to zero and a subsequent growth with distinct rates in each parametrization. This behavior on Q$^{2}$ occurs due to the special form of the function ${\cal{X}}(\boldsymbol{k}^{2},Q^{2})$, which diverges for Q$^{2} =$ 0.

\begin{figure}[t!]
\centering
\resizebox{\textwidth}{!}{
\rotatebox{-90}{\scalebox{00.50}{\includegraphics*[83pt,25pt][555pt,723pt]{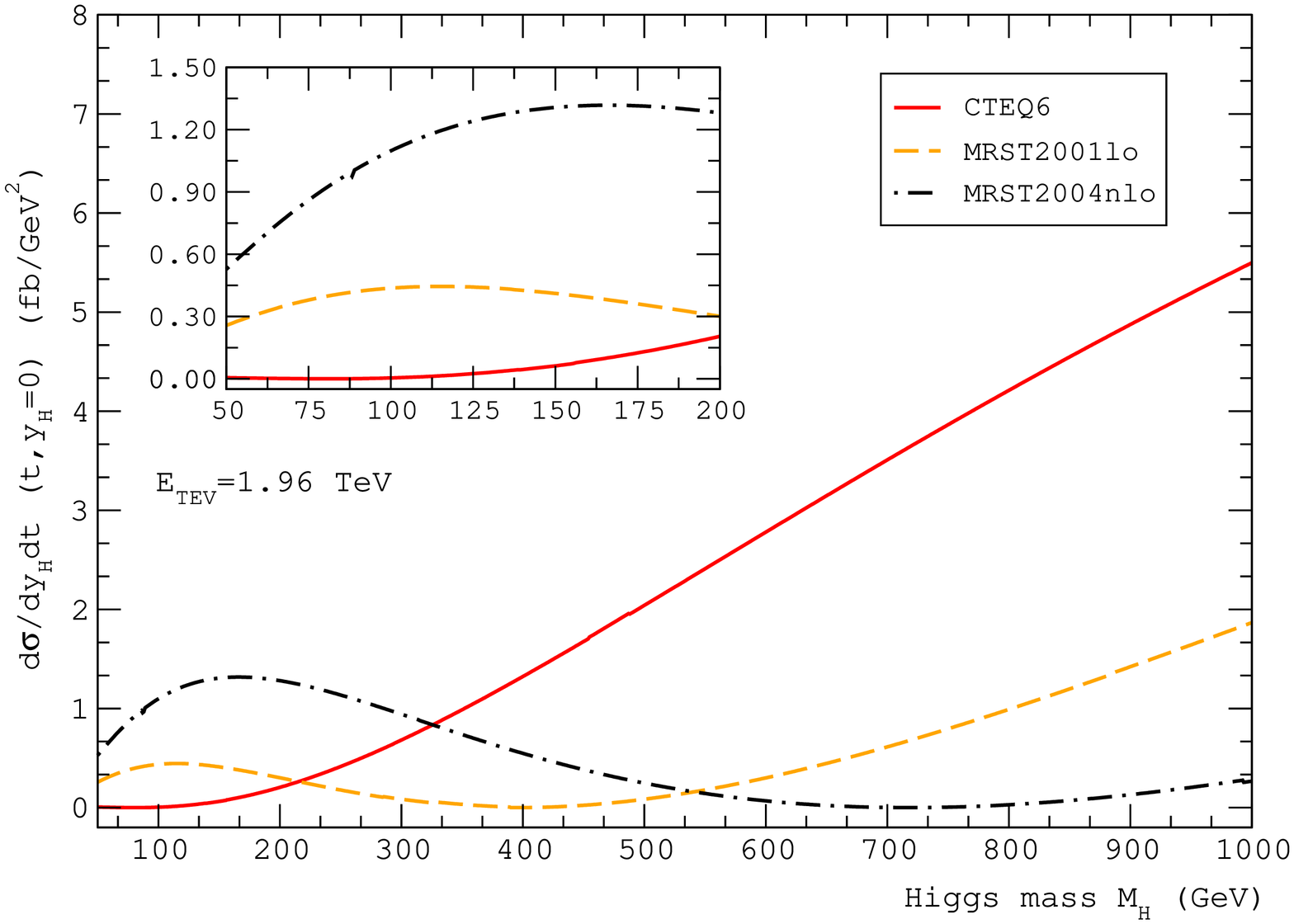}}}
\hfill
\rotatebox{-90}{\scalebox{00.50}{\includegraphics*[83pt,25pt][555pt,723pt]{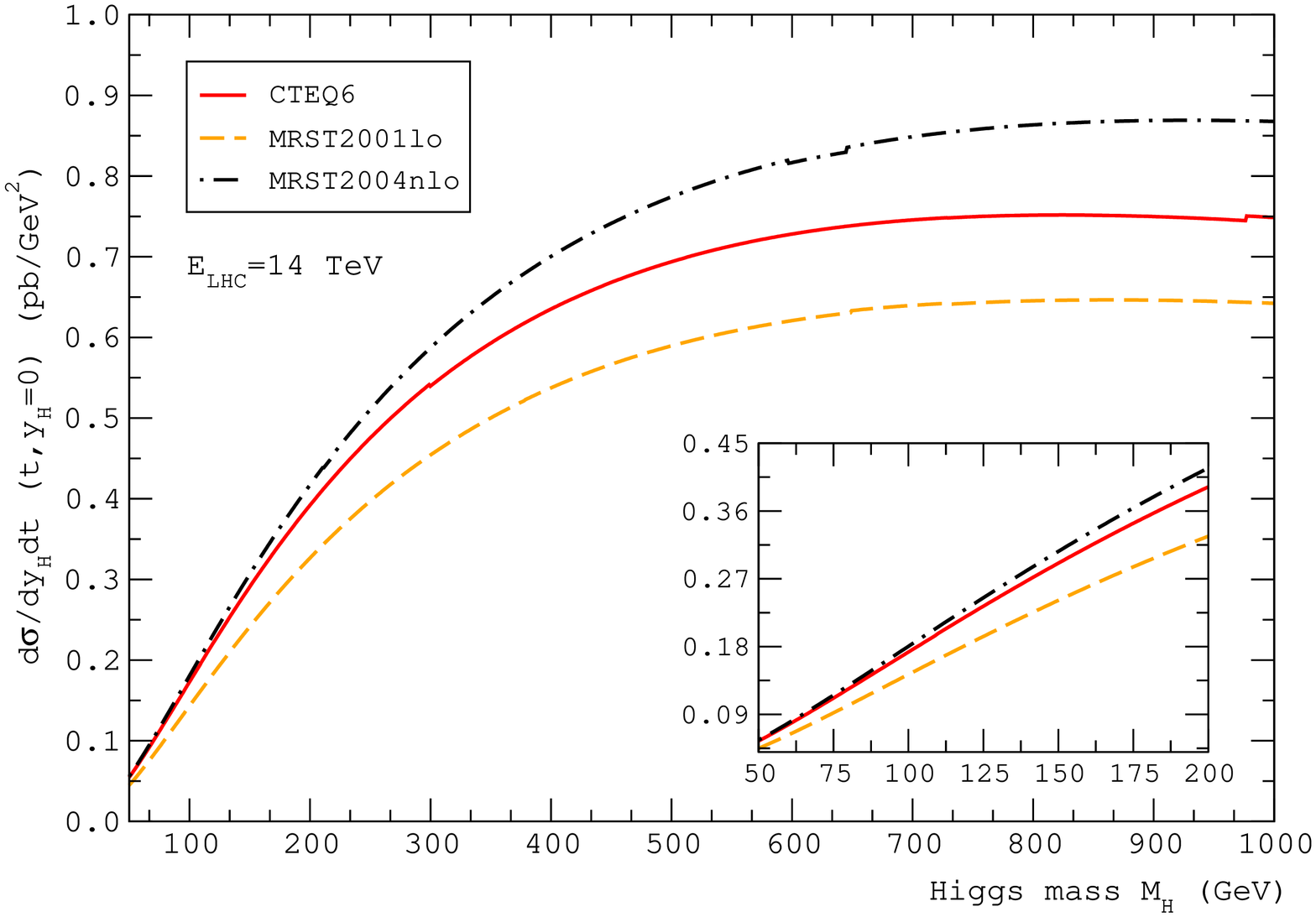}}}
}
\vskip1pc
\caption{\label{fig:fig5} Event rate $d\sigma/dy_{H}$ ($y_{H}$=0) in function of the Higgs mass for energy of Tevatron (left) and LHC (right), taking the results in an intermediary mass range (smaller graph) and an extended one (bigger graph), describing the effect of using distinct parametrizations for the gluon distribution functions. The results for Tevatron are shown in $fb$, the LHC ones are in $pb$.}
\end{figure}

Extending this numerical analysis, the event rate is predicted adopting some distribution functions for the gluon content in the proton. As explained in the Sec. III, the non-diagonality of the distributions was approximated by a multiplicative factor which permits one to employ the usual diagonal distributions. Consequently, the event rate is computed using one LO distribution and two distinct Next-to-Leading Order (NLO) distributions: MRST2004 and CTEQ6. This second possibility expresses our intention to analyse the impact of the gluon recombination effects in Higgs production at LHC. All these distributions were evolved from an initial momentum $\boldsymbol{k}_{0}^{2} = 1.0\textrm{ GeV}^{2}$, assumed as a mean-value between the initial scales for each parametrization. The results are shown in Fig.~\ref{fig:fig5} for two different mass ranges, and taking predictions for Tevatron and LHC energies. In the result for LHC, the event rate has a different shape in comparison to the results for Tevatron in virtue of the energy scale. The momentum fraction in Tevatron does not reach the necessary value $x = 0.01$ to permit one to consider the coupling of a gluon ladder to the proton. At most, the momentum fraction in Tevatron achieves $x \approx$ 0.05, when considering the lower bound of the Higgs mass (114.4 GeV \cite{lep}), which is not enough to employ the unintegrated gluon distribution. This is an important result and reveals the limitations of this approach. Otherwise, this kind of discrepancy is not observed in LHC, since the necessary value of momentum fraction can be easily achieved. Another important feature observed in LHC is the difference between the LO and NLO distributions. In this energy scale, the contributions from the recombination effects take place and reveal its importance to correctly predict the cross sections. Having a smaller energy compared to LHC, the Tevatron do not show the same evidence. Further information about the Higgs production and better knowledge on the recombination effects in QCD should be obtained in the future data from LHC. As made before in the LHC predictions, the mass range is extended in order to observe the role of the Sudakov form factors, showing the same behavior for all parametrizations.

\begin{figure}[t!]
\centering
\resizebox{\textwidth}{!}{
\rotatebox{-90}{\scalebox{00.50}{\includegraphics*[80pt,25pt][585pt,730pt]{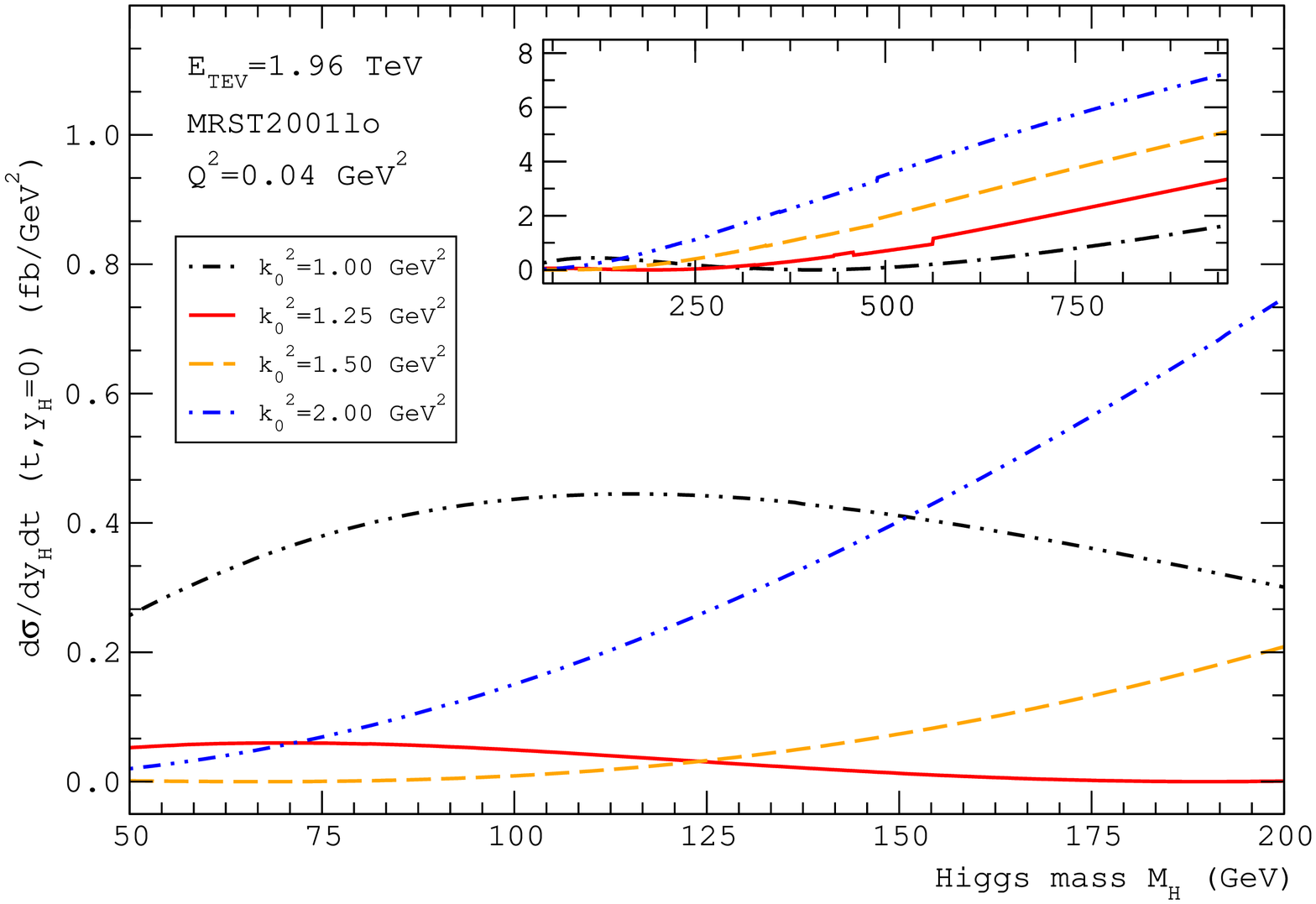}}}
\hfill
\rotatebox{-90}{\scalebox{00.485}{\includegraphics*[80pt,25pt][590pt,730pt]{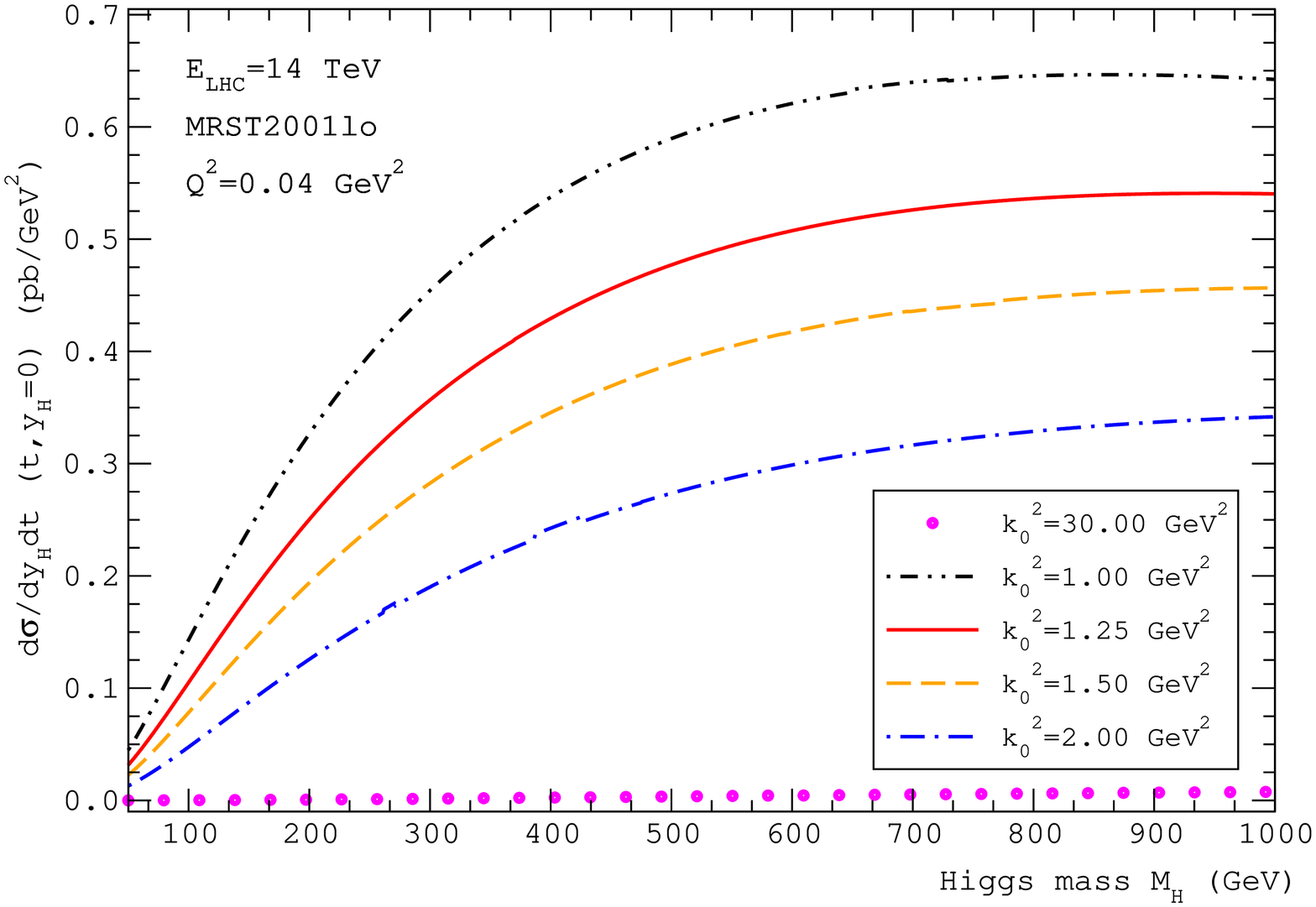}}}
}
\caption{\label{fig:fig6} Event rate $d\sigma/dy_{H}$ ($y_{H}$=0) in function of the Higgs mass. The graphs on the left-hand side show the results obtained considering cuts of the momentum integration for Tevatron energy, where the contribution for the event rate is observed in two ranges of the Higgs mass. The graph in the right-hand side shows the results for LHC energy (the results for the Tevatron are shown in $fb$ while for LHC are shown in $pb$).}
\end{figure}

An analysis must be done to verify the sensitivity of the results to the cut in the momentum integration, for that the event rate is calculated for some values of the cut, as shown in Fig.~\ref{fig:fig6}. There is a significative difference between the results of Tevatron and LHC, where the contributions to the event rate have very distinct behaviors as the cut value varies. This comparison shows that there is a smaller contribution to the event rate in LHC for higher cut values, being almost two times smaller if one takes $\boldsymbol{k}^{2}_{0} = 2.0\textrm{ GeV}^{2}$, instead of $\boldsymbol{k}^{2}_{0} = 1.0\textrm{ GeV}^{2}$. On the other hand, its contribution in Tevatron is quite distinct than the LHC results. Analysing it in the intermediary mass range, there is a non-uniform behavior between the results of Tevatron, where, roughly speaking, the results with the extreme values have a leading contribution for the event rate. However, the results obtained in the extended mass range show that for higher values of the Higgs mass the contribution is exactly contrary than those obtained for LHC, differing in four times the results between the extreme values. This behavior in the results for Tevatron shows the consequence of taking a higher value to the momentum fraction ($x \sim 0.05$). Therefore, analysing the results for these two energy scales, the sensitivity is as was expected, showing a small variation for different cut values, e.g., for a Higgs mass of 140 GeV, these results are three times less sensitive than the KMR ones. In the results for LHC, one can estimate an upper limit for the momentum cut, such that the contributions to the event rate can be neglected from this cut value: the result for $\boldsymbol{k}^{2}_{0}$ = 30 GeV$^{2}$ is almost zero in all range. This sensitivity can be explained by the form of the differential cross section obtained from this approach if compared with the result of the Durham group, where there is a higher sensitivity. Despite of an additional distribution function, in the photoproduction approach the event rate has a dependence on $\boldsymbol{k}^{-6}$ and a function depending on $Q^{2}$ and $\boldsymbol{k}^{2}$.

\begin{figure}[t!]
\centering
\resizebox{\textwidth}{!}{
\rotatebox{-90}{\scalebox{01.00}{\includegraphics*[87pt,16pt][570pt,718pt]{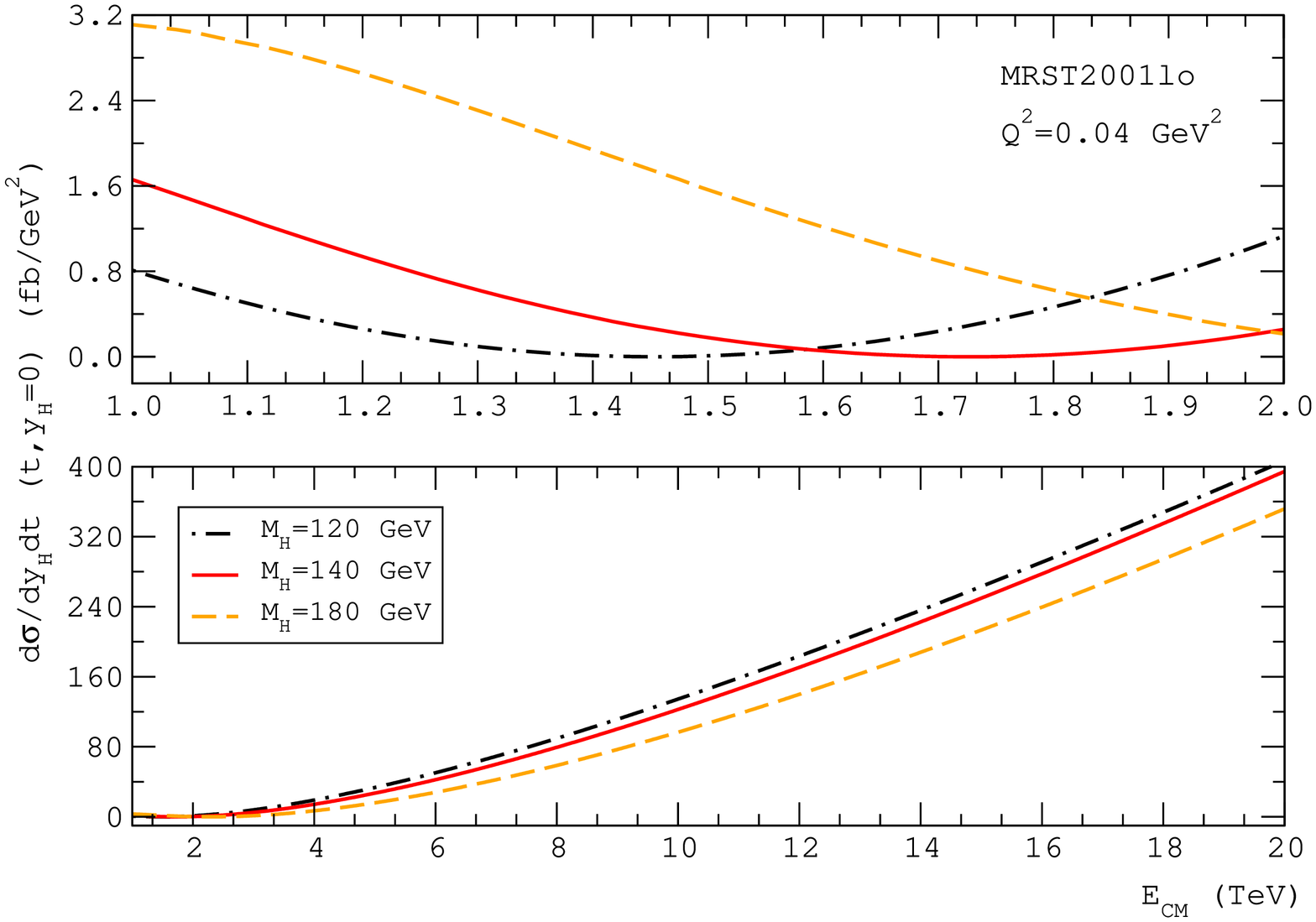}}}
\hfill 
\rotatebox{-90}{\scalebox{01.00}{\includegraphics*[87pt,15pt][551pt,718pt]{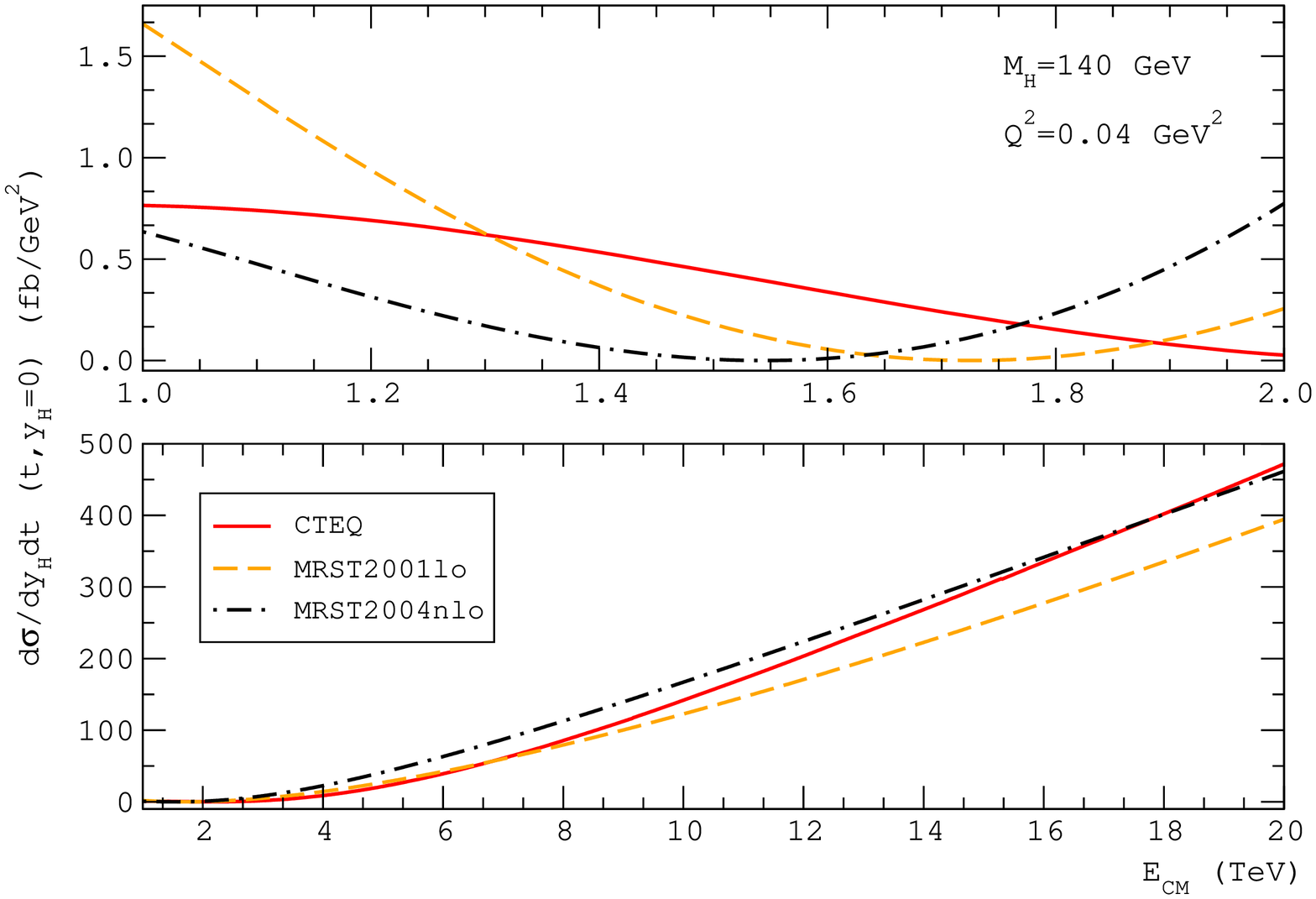}}}
}
\caption{\label{fig8} Event rate $d\sigma/dy_{H} (y_{H} = 0)$ in function of the center-of-mass energy of the process. The results on the left-hand side assume some values to the Higgs boson mass using the parametrization MRST2001. The right-hand ones are obtained with the same set of parametrizations used before and a Higgs mass of $140\textrm{ GeV}$.}
\end{figure}

To observe the dependence of the event rate on the center-of-mass energy, the results are obtained in function of the E$_{\textrm{CM}}$ for distinct Higgs masses and observing the behavior with the distribution functions. This process is analysed for three values of the Higgs boson mass, as shown in Fig.~\ref{fig8}. As can be seen, the growth of the event rate with E$_{\textrm{CM}} = \sqrt{s}$ has a parabolic shape up to E$_{\textrm{CM}}$ = 2.0 TeV (Tevatron region) and a linear behavior for higher energies. As explained in this Section, this parabolic shape occurs due to the values of $x$ probed in this energy region, showing approximately the same result for any chosen value of the Higgs boson mass. In the upper graphs in Fig.~\ref{fig8}, one observes the event rate in a lower range of energy. As the Higgs mass varies, the results show a non-uniform behavior if compared to the one observed in the high-energy region: the contribution from $M_{H}$ = 120 GeV is smaller than the other ones, and grows as the energy increases, assuming a leading contribution in the very-high-energy limit. The dependence of the $x$-variable on the center-of-mass energy is the reason to the presence of this transition region. Morevoer, the Sudakov form factors determine the lower contribution to the production of a heavier Higgs boson. The same behavior occurs in the results using distinct parametrizations, as shown in the upper-right graph in Fig.~\ref{fig8}. A similar contribution arises among the curves in the left-upper and right-upper graphs, although very distinct absolute values of the event rate. An important aspect observed in the lower-right graph is the same difference shown between MRST LO and NLO distribution functions, as seen in all energy range. The CTEQ6 parametrization has a transition behavior: this is similar to MRST2001 up to E$_{\textrm{CM}}$ = 8.0 TeV, and then grows to achieve MRST2004 at very-high energy.

\section{DISCUSSION}\label{sec:disc}

The contribution to the Higgs production for distinct virtualities shows a dependence upon the photon energy. Observing the comparison with the results of \cite{forshaw-KMR}, the photoproduction results have a higher contribution in the intermediary range as well as in the extended one. However, when the full analysis in Peripheral Collisions is taken into account, including the photon distribution in the proton, this difference between the approaches is expected to be modified, in order to the photoproduction results achieve the same shape of those of the Durham group, although with a higher contribution in the intermediary range. Otherwise, the study of this process in nucleus-nucleus collisions will show its dependence on the photon energy and on the photon number, features that should be drastically modified if compared to the $pp$ case. As one can see, the results obtained taking the photon virtuality on the order of Q$^{2} \lesssim$ 10$^{-2}$ GeV$^{2}$ gives an event rate going to infinity, such that the real-photon limit is reached. Thus, even considering an overestimated prediction to the Higgs production at small virtualities, the results at Q$^{2}$ = 1.0 GeV$^{2}$ reach values of a few femtobarns, which agree with other predictions for diffractive Higgs production at LHC \cite{KMR,miller,levin}. Waiting the data coming from CMS and ATLAS experiments at LHC, these results will be confronted to the data in order to determine the best options for Higgs photoproduction in Peripheral Collisions. One of the main aspects to be observed is to get suitable data to specify the restrictions to the virtuality range.

\section{CONCLUSIONS}\label{sec:ccl}

A new way to produce the Higgs boson was studied in Peripheral Collisions, calculating perturbatively the event rate for diffractive production through DPE. Previously, some studies had been done exploring the photoproduction process, however none of them adopting the DPE as the interaction between the colliding particles. The numerical results obtained from the photoproduction approach predicts a reasonable event rate for Higgs production at LHC if compared to previous estimates for the Higgs production. The event rate was obtained for the $\gamma p$ interaction with a dependence on $\boldsymbol{k}^{-6}$, unlike to the result carried out in \cite{forshaw-KMR}. To effectively compare the results presented in this work with those of the Durham group, a distribution function for the photons in the proton should be introduced, and then the results to the peripheral $pp$ collisions will be computed. Therefore, the results show the possibility to produce the Higgs boson through Peripheral Collisions at LHC with an event rate expected to be big enough to detect this boson.

\section{ACKNOWLEDGEMENTS}
GGS would like to thank W. Sauter for discussions in numerics, MBGD acknowledges the hospitality of CBPF during the completion of this work and enlightening discussions with Maria Elena Pol, Ronald Shellard, Alberto Santoro and Uri Maor. This work was partially supported by CNPq (GGS and MBGD).


\begin{thebibliography}{99}

\bibitem{KMR-1997} V.A. Khoze, A.D. Martin, M.G. Ryskin, Phys. Lett. B \textbf{401}, 330 (1997).

\bibitem{bauretal} G. Baur et al, Phys. Rept. \textbf{364} 359 (2002).

\bibitem{bertu} C.A. Bertulani, Heavy Ion Phys. \textbf{14} 51 (2001) 

\bibitem{hencken} K. Hencken et al, Phys. Rept. \textbf{458} 1 (2008).

\bibitem{forshaw-book} J.R. Forshaw, D.A. Ross, \textit{Quantum chromodynamics and the pomeron} (Cambridge University Press, Cambridge, 1997).

\bibitem{barone} V. Barone, E. Predazzi, \textit{High-Energy Particle Diffraction} (Springer-Verlag, Berlin, 2002).

\bibitem{frankfurt} L.L. Frankfurt, A. Freund, M. Strikman, Phys. Rev. D \textbf{58} 114001 (1998).

\bibitem{kuchina} I. Balitsky, E. Kuchina, Phys. Rev. D \textbf{62} 074004 (2000).

\bibitem{bialas} A. Bialas, P.V. Landshoff, Phys. Lett. B \textbf{256} 540 (1991).

\bibitem{forshaw-evanson} N.G. Evanson, J.R. Forshaw, Phys. Rev. D \textbf{60} 034016 (1999).

\bibitem{foldy} L. Foldy, R.F. Peierls, Phys. Rev. \textbf{130} 1585 (1963).

\bibitem{mueller1} A.H. Mueller, Nucl. Phys. B \textbf{415} 373 (1994); Nucl. Phys. B \textbf{437} 107 (1995).

\bibitem{fORM} J.A.M.Vermaseren, arXiv:math-ph/0010025.

\bibitem{kniehl} B.A. Kniehl, Phys. Rep. \textbf{240} 211 (1994).

\bibitem{forshaw-KMR} J.R. Forshaw, arXiv:hep-ph/0508274.

\bibitem{KMR-CanThe} V.A. Khoze, A.D. Martin, M.G. Ryskin, Eur. Phys. J. C \textbf{14} 525 (2000).

\bibitem{off-diag} K.J. Golec-Biernat, A.D. Martin, Phys. Rev. D \textbf{59} 014029 (1998).

\bibitem{miller} J.S. Miller, arXiv:0704.1985[hep-ph].

\bibitem{KMR-Rates} V.A. Khoze, A.D. Martin, M.G. Ryskin, arXiv:hep-ph/0103007.

\bibitem{shuvaevetal} A.G. Shuvaev, K.J. Golec-Biernat, A.D. Martin, M.G. Ryskin, Phys. Rev. D \textbf{60} 014015 (1999).

\bibitem{bjorken1993} J.D. Bjorken, Phys. Rev. D \textbf{D47} 101 (1993).

\bibitem{chehime} H. Chehime et al, Phys. Lett. B \textbf{286} 397 (1992).

\bibitem{KMR} V.A. Khoze, A.D. Martin, M.G. Ryskin, Eur. Phys. J. C \textbf{18} 167 (2000).

\bibitem{luna} E.G.S. Luna, Phys. Lett. B \textbf{641} 171 (2006).

\bibitem{gotsman} E. Gotsman, E. Levin, U. Maor, Phys. Lett. B \textbf{438} 229 (1998); Phys. Rev. D \textbf{60} 094011 (1999).

\bibitem{GRV} M. Gl\"uck, E. Reya, A. Vogt, Nucl. Phys. B \textbf{130} 76 (1977); Z. Phys. C \textbf{67} 433 (1995); Eur. Phys. J. C \textbf{5} 461 (1998).

\bibitem{MRST} A.D. Martin et al, Eur. Phys. J. C \textbf{4} 463 (1998); Eur. Phys. J. C \textbf{14} 133 (2000); Phys. Lett. B \textbf{531} 216 (2002).

\bibitem{ALEKHIN} S.I. Alekhin, Phys. Rev. D \textbf{68} 014002 (2003).

\bibitem{CTEQ} J. Pumplin et al, JHEP \textbf{0207} 012 (2002).

\bibitem{lhapdf} LHAPDF project, arXiv:hep-ph/0508110 $\med{\textrm{http://hepforge.cedar.ac.uk/lhapdf/}}$.

\bibitem{lep} R. Barate et al, Phys. Lett. B \textbf{565} 61 (2003).

\bibitem{levin} E. Levin, J.S. Miller, arXiv:0801.3593[hep-ph].

\end{thebibliography}
\end{document}